\documentclass{optica-article}

\journal{opticajournal} 

\articletype{Research Article}

\usepackage{lineno}
\usepackage{siunitx}

\begin{document}

\title{2D Optical Beam Scanning using Integrated Acousto-Optics and a Frequency Comb}

\author{Shucheng Fang\authormark{1}, Qixuan Lin\authormark{1}, Fengyan Yang\authormark{3}, Yue Yu\authormark{1}, Guangcanlan Yang\authormark{3}, Bingzhao Li\authormark{1}, Hong X. Tang\authormark{3}, and Mo Li\authormark{1,2,*}}

\address{\authormark{1} Department of Electrical and Computer Engineering, University of Washington, Seattle, WA 98115, USA\\
\authormark{2} Department of Physics, University of Washington, Seattle, WA 98115, USA\\
\authormark{3} Department of Electrical Engineering, Yale University, New Haven, Connecticut 06511, USA}

\email{\authormark{*}moli96@uw.edu} 


\begin{abstract*} 
Optical beam steering is an essential technology for free-space optical communication, reconfigurable optical networks and quantum information systems. Yet conventional steering methods either require bulky mechanical mechanisms, or rely on complex arrays of individually controlled light emitting elements. Integrated acousto-optic beam steering (AOBS) offers non-mechanical, continuous one-dimensional steering on-chip by using traveling acoustic waves with variable frequency to deflect light. In this work, we combine AOBS with an optical frequency comb and optical gratings to enable two-dimensional beam steering from a single aperture. Azimuthal scanning is controlled via acoustic frequency while polar coverage is realized by dispersing frequency comb lines with the gratings. We demonstrate this architecture by sequentially selecting and steering 11 comb lines spanning 1540–1570 nm, achieving a field of view of \SI{18.2}{\degree} $\times$ \SI{4.3}{\degree}. Validation with a tunable laser extends polar coverage to \SI{11.4}{\degree}. Both components are realized on the same thin-film lithium niobate platform, providing a pathway toward monolithic integration.
\end{abstract*}

\section{Introduction}

Solid-state beam steering enables dynamic spatial control of optical fields for many emerging applications, including free-space optical communication (FSOC) \cite{rabinovich2015free, he2020review}, reconfigurable optical networks \cite{liverman2023dynamic, liu2023hybrid}, and quantum information systems \cite{bluvstein2024logical}. In FSOC, for example, agile beam steering is desired for point-to-point links to track moving platforms and establish communication channels without mechanical gimbals. Many technologies have been developed to realize non-mechanical beam steering, including optical phased arrays (OPAs) \cite{liu2022silicon}, focal plane switch arrays (FPSAs) \cite{zhang2022large}, and spatial light modulators (SLMs) \cite{zhang2024scaled}. All of them use apertures synthesized by a large number of individually controlled light emitting elements, each requiring phase or amplitude controls with corresponding control circuitry, which leads to challenges in scalability.

Integrated acousto-optic beam steering (AOBS) offers a compelling alternative by using traveling acoustic waves to deflect light through dynamically generated index gratings on-chip. The deflection angle is controlled by the acoustic frequency, enabling continuous one-dimensional (1D) scanning from a single aperture using only one acoustic transducer. While bulk acousto-optic devices \cite{trypogeorgos2013precise} have been used for decades, integrated platforms enable much higher optical and acoustic power densities, dramatically improving efficiency, compactness, and steering field of view (FOV). Our previous work \cite{li2023frequency, lin2024optical} has successfully demonstrated the technology and its application in LiDAR and FSOC.

To achieve two-dimensional steering, wavelength-dependent diffraction is a compelling approach. By combining beam steering along one axis with wavelength-dependent diffraction along another, 2D FOV coverage can be achieved. This approach has been explored with tunable lasers in OPAs \cite{liu2022silicon} and, more recently, with optical frequency combs that provide multiple discrete wavelength channels simultaneously. Recent work demonstrated this concept by combining microcombs with OPAs \cite{chen2025single} for parallel LiDAR operation.

Here, we extend the AOBS architecture from 1D steering to 2D by combining it with an integrated Kerr frequency comb on the thin-film lithium niobate (TFLN) platform. The AOBS provides azimuthal scanning through acoustic frequency tuning, while grating dispersion separates the comb lines in the polar direction. We demonstrate this architecture with comb lines spanning 1540-1570 nm, achieving \SI{18.2}{\degree} $\times$ \SI{4.3}{\degree} 2D FOV from a single continuous aperture. Both the frequency comb and AOBS are realizable on the TFLN platform, providing a pathway toward monolithic integration.

\begin{figure}[ht]
    \centering
    \includegraphics[width=1\linewidth]{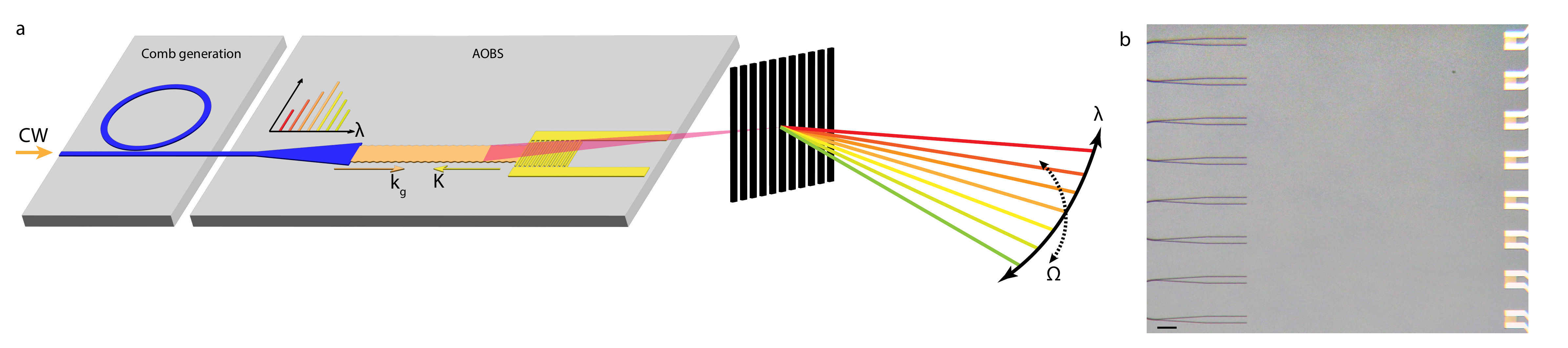}
    \caption{Acousto-optic beam steering of frequency comb. (a) Schematics of the system. A microring resonator is pumped by a CW laser to generate a comb on the first chip. The comb is then transmitted onto the second chip and steered into free space by the acousto-optic beam steering system, and then diffracted by another grating to generate 2D scanning pattern. $\lambda$ and $k_g$ are the optical wavelength and guided wavevector, $\Omega$ and $K$ are the acoustic frequency and wavevector. (b) Microscope image of the AOBS system. Scale bar: \SI{100}{\micro\meter}.}
    \label{fig:schematics}
\end{figure}

\section{Device design}

\subsection{Optical frequency comb}

\begin{figure}[ht]
    \centering
    \includegraphics[width=1\linewidth]{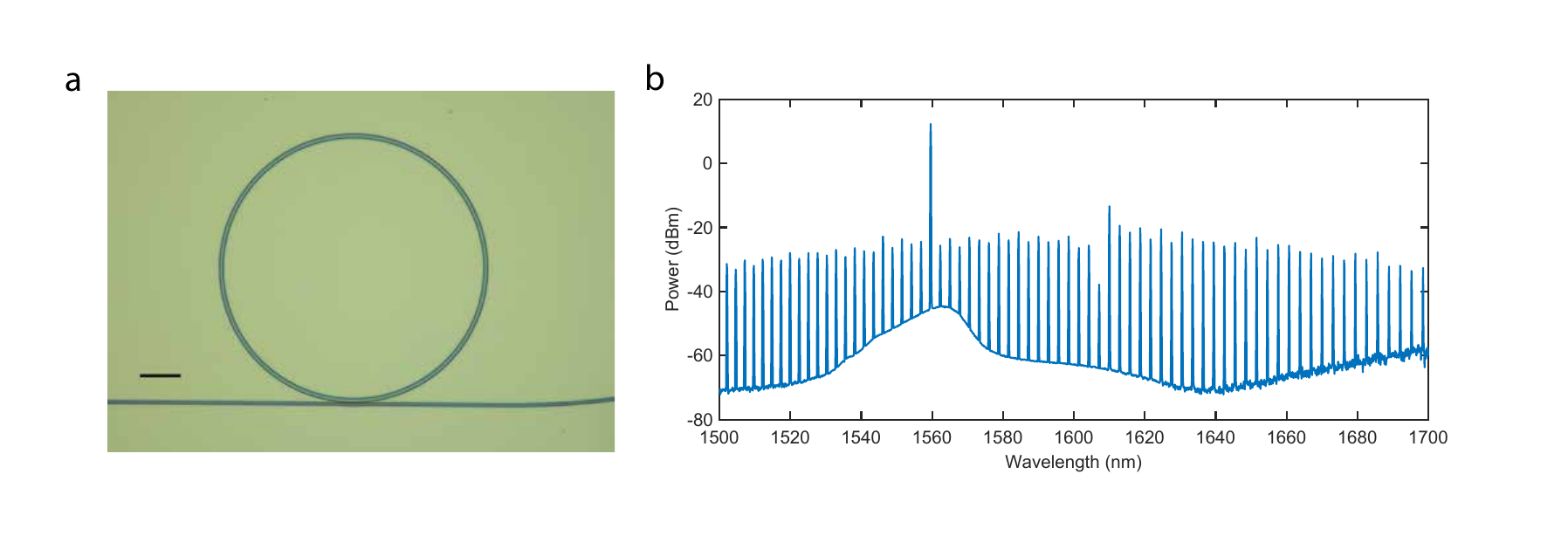}
    \caption{Characterization of the frequency comb. (a) Microscope image of the microring resonator. Scale bar: \SI{10}{\micro\meter}. (b) Spectrum of the generated soliton.}
    \label{fig:comb}
\end{figure}

The frequency comb is generated from a lithium niobate microring resonator fabricated on a 600-nm x-cut lithium-niobate-on-insulator (LNOI) wafer. As shown in Fig.~\ref{fig:comb}a, the ring has a radius of \SI{60}{\micro\meter} and a waveguide width of \SI{1.4}{\micro\meter} with a free spectral range of \SI{340}{\giga\hertz}, designed for anomalous group velocity dispersion to support dissipative Kerr soliton formation. The loaded quality factor is measured to be $6.04\times10^5$ in the transmission spectrum. 

When pumped with \SI{200}{\milli\watt} of continuous-wave light at \SI{1550}{nm} amplified by an erbium-doped fiber amplifier (EDFA), the microring generates a soliton. The output spectrum, measured with an optical spectrum analyzer (OSA), shows a soliton comb spanning >\SI{200}{nm} bandwidth from \SI{1500}{nm} to \SI{1700}{nm} (Fig.~\ref{fig:comb}b).

\subsection{Acousto-optic beam steering}

The AOBS device follows the design in Ref.~\cite{lin2024optical} and consists of three parts: an optical input coupler and waveguide, an interdigital transducer (IDT) for generating surface acoustic wave (SAW), and the AO interaction area where the optical and acoustic waves interact and the beam is deflected, as shown in Fig.~\ref{fig:schematics}b. The input light is coupled into a single mode waveguide via edge coupling and then slowly expanded into a \SI{30}{\micro\meter} wide slab mode in order to maximize the AO interacting region and improve beam profile while also reducing background scattering caused by etched waveguide edges. The IDTs on the other end of the AO aperture are designed with a linearly chirped period from \SI{1401}{nm} to \SI{1827}{nm} to achieve a higher bandwidth. The design utilizes Rayleigh mode surface acoustic waves for their high acousto-optic coupling efficiency.

The steering mechanism relies on Brillouin scattering between the guided optical mode and counter-propagating surface acoustic waves. The deflection angle $\theta$, which is the angle between the deflected beam and the chip surface plane, is determined by the phase-matching condition:
\[\sin\theta = \frac{k_g-K}{k_0} = n_{\text{eff}}-\frac{\Omega}{k_0v_{\text{SAW}}}\]
where $k_0$ and $k_g$ are the free-space and guided optical wavevector, $\Omega$ is the acoustic frequency, $n_{\text{eff}}$ is the effective refractive index, and $v_{\text{SAW}}$ is the SAW velocity in lithium niobate. By sweeping $\Omega$ from 1.8 to \SI{2.1}{\giga\hertz}, continuous azimuthal scanning over \SI{18}{\degree} field of view is achieved.

\subsection{Gratings}

For dispersing the comb lines, transmissive metal gratings with a pitch of \SI{880}{nm} are fabricated on a silicon chip. Combined with an incident angle of \SI{60}{\degree}, the gratings can achieve a FOV of \SI{4.7}{\degree} for 1540-\SI{1570}{nm} (EDFA bandwidth) and \SI{9.7}{\degree} for 1500-\SI{1570}{nm} (tunable laser bandwidth).

\section{Experimental Setup and Characterization Methods}

\begin{figure}[htbp]
    \centering
    \includegraphics[width=\linewidth]{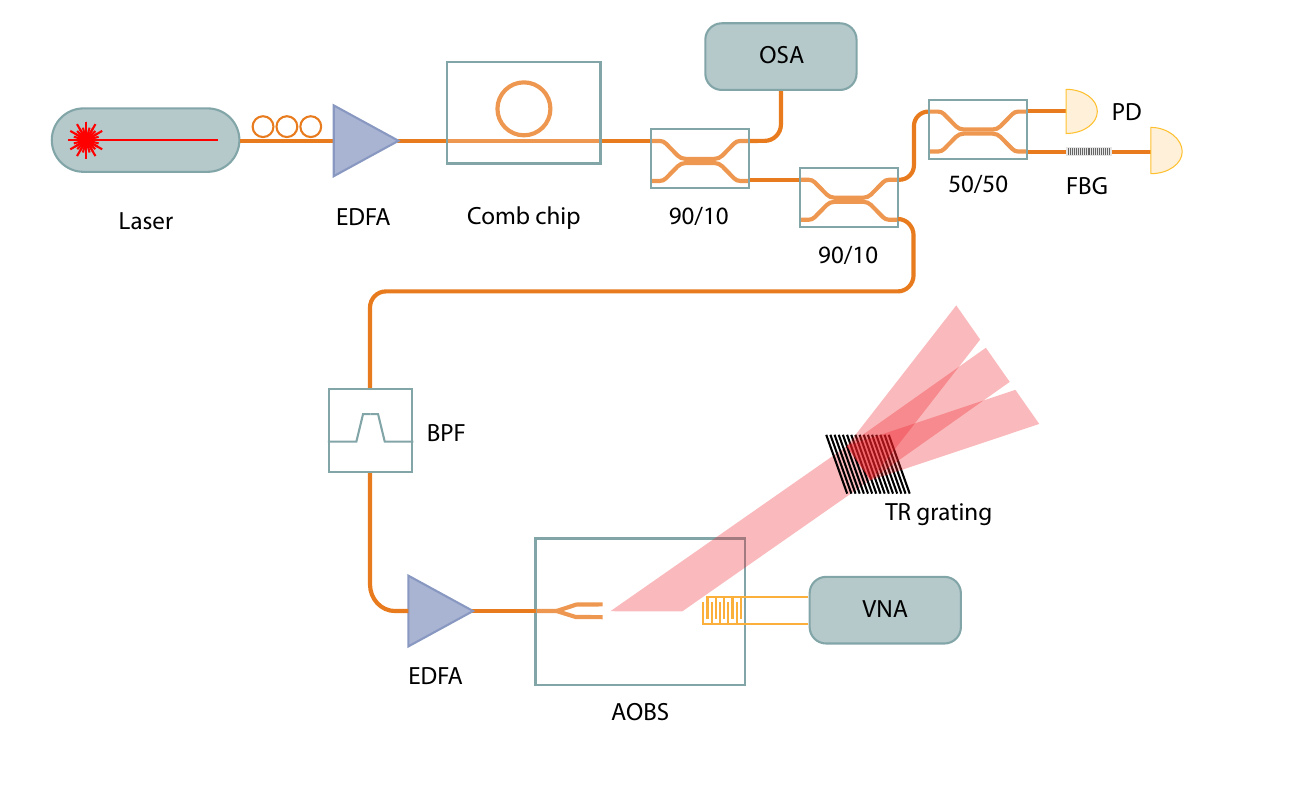}
    \caption{Experimental setup for 2D beam steering. A soliton frequency comb is generated from a lithium niobate microring resonator. Individual comb lines are selected using a tunable bandpass filter (BPF) and amplified via an erbium-doped fiber amplifier (EDFA) before coupling to the acousto-optic beam steering (AOBS) device. The AOBS chip, driven by RF signals from a vector network analyzer (VNA), steers the beam in the azimuthal direction. The steered beams pass through transmissive (TR) gratings providing wavelength dispersion in the polar direction. A portion of the comb output is monitored using an optical spectrum analyzer (OSA) and the comb power is monitored via fiber Bragg gratings (FBG).}
    \label{fig:system}
\end{figure}

In the experiment setup, the three components described above are assembled, as shown in Fig.~\ref{fig:system}. A \SI{1550}{nm} laser is amplified by the first EDFA to reach a power level of \SI{200}{mW}. It is then coupled to the comb chip via a fiber-lens edge coupler and the output is collected with a lensed fiber. Such design ensures stable coupling at high pump power. 

The output of the comb is split at the first 90/10 splitter, which directs 10\% of the output to an OSA and the remaining 90\% passes to a second 90/10 splitter. From the second splitter, 10\% is divided via a 50/50 coupler to two photodiodes: one monitors total transmission power, while the other measures comb power after filtering the residual pump via fiber Bragg gratings (FBG). The remaining output (81\% of initial power) proceeds to the AOBS.

The comb output first passes through a tunable bandpass filter (BPF) to select an individual comb line. It is then amplified by another low-noise EDFA and coupled into the AOBS chip. This approach ensures sufficient optical power per wavelength channel, as simultaneous amplification of all comb lines by the EDFA would result in insufficient signal strength for subsequent far-field beam characterization. The beam deflected from the chip is directed onto the transmissive gratings with an incident angle of \SI{60}{\degree}, and is collected by a k-space imaging system. Apart from the conventional setup, a cylindrical lens is included in the imaging system to compress the beam size in the polar direction, which is large due to the small acoustic aperture. The IDT is driven by the RF signal generated by a vector network analyzer (VNA).

The experiment starts by generating and stabilizing the comb. We acquire transmission and comb power spectrum across the entire EDFA bandwidth. Here, for consistency, we refer to the second photodiode measurement as the "comb power" even before comb generation. Peaks in the comb power spectrum corresponding to dips in the transmission spectrum indicate cavity resonances. One resonance with high comb power and high Q-factor is selected and swept repeatedly to achieve thermal stability. The pump wavelength is then fine-tuned until soliton steps appear on the comb power monitor and a characteristic soliton spectrum is observed on the OSA.

With the comb stabilized, we proceed to beam steering characterization. We first extract comb line positions from the OSA spectrum and scan the tunable bandpass filter across a small wavelength range around each peak to find the optimal transmission setting. This compensates for potential non-uniformity in the filter response. Individual comb lines are then selected and amplified via the EDFA. For each selected wavelength, the VNA sweeps the RF frequency from 1.8 to \SI{2.1}{\giga\hertz} in \SI{10}{\mega\hertz} steps. At each frequency step, the camera captures the far-field beam pattern with the RF signal on (signal image) and off (background image). Background subtraction is conducted during post-processing to remove ambient light and detector noise. This process is repeated for 11 comb lines spanning from 1540 to \SI{1570}{nm}, yielding a two-dimensional map of beam positions as a function of acoustic frequency and optical wavelength.

\section{Result and Discussion}

\begin{figure}[ht]
    \centering
    \includegraphics[width=1\linewidth]{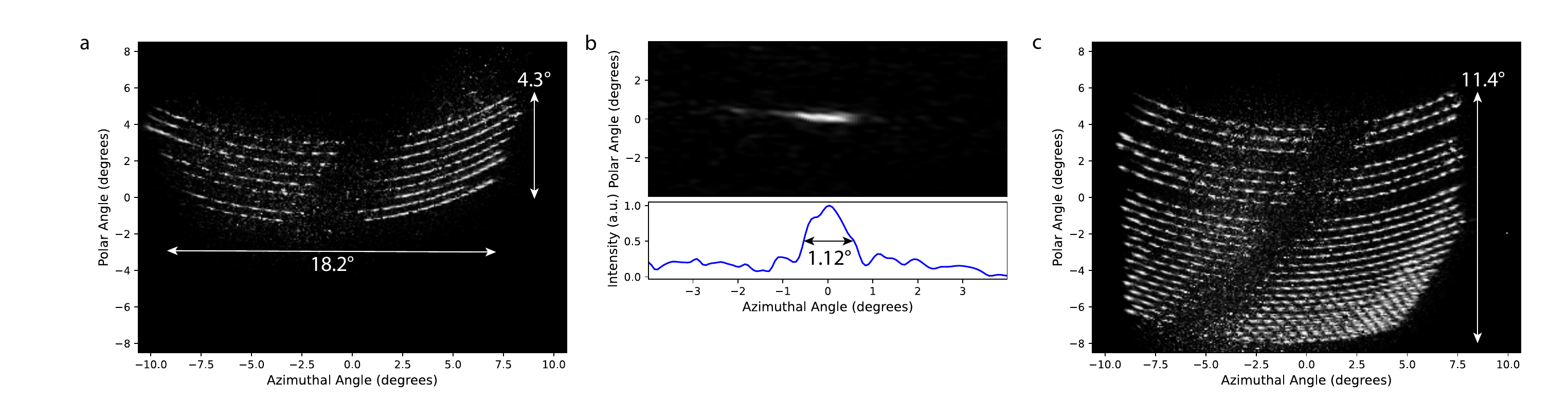}
    \caption{Two-dimensional beam steering pattern from k-space measurement of the system. (a) Superimposed k-space image from measurements at 11 comb lines and 31 RF frequencies. We were able to obtain a total FOV of \SI{18.2}{\degree} $\times$ \SI{4.3}{\degree}. (b) Image of a single beam spot. It has a \SI{1.12}{\degree} $\times$ \SI{0.40}{\degree} size and an on-off ratio of \SI{9.65}{\decibel}. The lower panel shows the horizontal cross-section of the image at the center of the beam spot. (c) Superimposed k-space image with the tunable laser. The FOV is expanded to \SI{18.2}{\degree} $\times$ \SI{11.4}{\degree}.}
    \label{fig:result}
\end{figure}

The k-space image shown in Fig.~\ref{fig:result}a is superimposed from measurements at 11 sequentially selected comb lines and 31 RF frequencies. Overall we were able to achieve a FOV of \SI{18.2}{\degree} in the azimuthal direction and \SI{4.3}{\degree} in the polar direction within the EDFA bandwidth, which agrees with previous results. 

Each comb line forms an arc shape instead of a straight line because of the transmissive gratings, which, while placed along the polar direction, also diffract in the azimuthal direction. The AOBS itself also has a dispersive effect on the comb lines, causing the arcs to shift horizontally, which is more obvious in Fig.~\ref{fig:result}c. While this represents a deviation from ideal rectilinear scanning, it does not fundamentally limit the 2D coverage capability. The low-intensity region in the center of the pattern is caused by non-uniform IDT response, which can be improved through optimized transducer design or power normalization \cite{morgan2010surface}. 

For each beam spot, background subtraction and spatial Gaussian filtering are applied to reduce background scattering. Due to non-uniform response of the IDT and EDFA, beam spots with insufficient deflected power are filtered out based on a peak intensity threshold. For the remaining spots, the beam position at each acoustic frequency is identified as the peak-intensity pixel after background subtraction. We utilize the random sample consensus (RANSAC) algorithm to fit the beam positions at the same optical wavelength to an arc, and exclude the points marked as outliers. The excluded spots predominantly occur at wavelength \SI{1565.5}{nm} and \SI{1568.3}{nm} which are at the border of the EDFA bandwidth and in the low-intensity region in the center due to low IDT efficiency. This distribution confirms that the measurement failures arise from device efficiency limitations rather than fundamental steering mechanism issues.

To validate performance beyond the EDFA bandwidth limitation, we replace the comb source with a tunable laser source with a \SI{70}{nm} bandwidth. As shown in Fig.~\ref{fig:result}c, the FOV is expanded to \SI{18.2}{\degree} $\times$ \SI{11.4}{\degree}, demonstrating that the system bandwidth is limited by amplification bandwidth rather than fundamental device constraints. With broadband amplification or higher comb output power, the full \SI{200}{nm} comb bandwidth could potentially provide >\SI{40}{\degree} polar coverage. This also verifies the AOBS azimuthal steering performance across the extended wavelength range. The consistent azimuthal FOV at all wavelengths confirms the broadband nature of the acousto-optic interaction and the chirped IDT design.

Fig.~\ref{fig:result}b shows the zoomed-in image of a single beam spot at optical wavelength of \SI{1530.3}{nm} and acoustic frequency of \SI{2.07}{GHz}. The measured beam size is \SI{1.12}{\degree} $\times$ \SI{0.40}{\degree} (FWHM). The aforementioned cylindrical lens compressed the polar beam size, while the azimuthal one is still limited by acoustic aperture size. The measured on-off ratio of \SI{9.65}{\decibel} indicates moderate background suppression. The background consists of scattering from the AOBS chip, grating diffraction into non-primary orders, and ambient light.

The current result, however, does not represent the system's theoretical limits. We analyze the achievable performance by component optimization and architectural innovation. The AOBS FOV is directly determined by the IDT bandwidth, which is \SI{300}{MHz} for the current system. The limitation arises from multiple factors, including impedance matching, RF power handling, acoustic aperture width, transduction efficiency and bandwidth. Generally impedance matching, RF power handling and transduction efficiency impact the generated acoustic wave power density, which has direct influence on scattering efficiency. Acoustic aperture width affects the beam profile quality while bandwidth determines the FOV. While the demonstrated system prioritizes proof-of-concept, future work focusing on optimization using strategies such as split-finger design \cite{morgan2010surface} and on-chip impedance-matching network \cite{hugot2026approaching} would increase the bandwidth without compromising other metrics. If we assume a relative bandwidth of 25\% as demonstrated in \cite{stettler2025suspended}, the azimuthal FOV would be increased from \SI{18.2}{\degree} to over \SI{30}{\degree}. 

The solution demonstrated in this work sequentially selects and amplifies comb lines to ensure sufficient power per channel. However, this architecture inherently supports simultaneous multi-beam and multi-wavelength operation, which is the unique advantage of the frequency comb approach, unlocking applications in parallel ranging and wavelength-multiplexed communication. The frequency comb used is a lithium-niobate-based Kerr comb with a span of \SI{200}{nm} from 1500 to \SI{1700}{nm} and comb line power around \SI{-30}{dBm}. To utilize the full \SI{200}{nm} spectrum, the optical amplifier between the comb and AOBS must be removed, as no amplifier can cover such a large bandwidth. Therefore, a high-power comb would be necessary. With high power combs of \SI{-10}{dBm} per line, we can achieve \SI{20}{\micro\watt} of steered power per channel with an AOBS efficiency of 20\% \cite{yu2026resonance}. With the \SI{200}{nm} span, we can also achieve a FOV of \SI{33}{\degree}.

The current system incorporates free-space transmissive gratings for dispersing comb lines, which compromises the system's compactness and robustness. There are two possible solutions for integrating the gratings onto the chip. The first is to place the grating between the optical taper and the IDT, so that the optical beam is deflected in the polar direction first, and then goes into the AO aperture to be scattered out-of-plane. This design will be easier to fabricate, but reduces efficiency due to imperfect alignment between optical and acoustic waves. The second approach is to utilize the part of the AOBS beam that is deflected downward into the substrate instead of air, and fabricate immersion gratings \cite{van2017state} at the back of the chip. They can function similarly to the free-space grating while offering a larger dispersion and resolution due to the high refractive index \cite{tang2018dispersion}, but have a smaller design space as the incident angle is fixed. It can also be expanded into a metasurface for potential beam profile optimization.
 
\section{Conclusion}

We successfully demonstrated 2D beam steering by combining integrated AOBS with an on-chip frequency comb on thin-film lithium niobate. The system achieves \SI{18.2}{\degree} $\times$ \SI{4.3}{\degree} coverage through azimuthal scanning via RF frequency modulation and polar wavelength dispersion via gratings. Sequential characterization of comb lines validates the 2D steering concept, while tunable laser measurements confirm potential for >\SI{11}{\degree} polar range with broadband amplification. With future work such as IDT optimization, a FOV of over \SI{30}{\degree} $\times$ \SI{30}{\degree} is projected. The demonstrated architecture establishes a pathway toward compact 2D beam steering that avoids the scaling challenges of discrete element arrays while leveraging the natural multi-wavelength capability of integrated frequency combs.

Beyond beam steering, the acousto-optic platform demonstrates broader potential for programmable spatial light control through frequency-domain encoding. The demonstrated approach of mapping acoustic frequency to spatial beam position represents a broader paradigm that enables diverse spatial control implementations. While this work uses sinusoidal acoustic waves for beam deflection, complex RF waveform modulation can generate arbitrary acoustic field patterns, enabling programmable phase mask generation analogous to spatial light modulators. Operating at GHz acoustic frequencies, this approach could achieve MHz-rate reconfigurable phase modulation significantly faster than conventional SLM technologies \cite{jabbari2024fast}. The frequency comb can also be extended to generate arbitrary frequency combinations with frequency modulators, enabling spatial modulation on the polar direction as well \cite{wei202610}. Combining these techniques along orthogonal dimensions, analogous to the demonstrated 2D beam steering architecture, could enable 2D programmable phase masks.

\begin{backmatter}
\bmsection{Funding}
DARPA MTO SOAR program (Award No. HR0011363032)

\bmsection{Acknowledgments}
Part of this work was conducted at the Washington Nanofabrication Facility and Molecular Analysis Facility, a National Nanotechnology Coordinated Infrastructure (NNCI) site at the University of Washington with partial support from the National Science Foundation via award nos. NNCI-2025489. 

\bmsection{Disclosures}
The authors declare no conflicts of interest.

\bmsection{Data availability} Data underlying the results presented in this paper are not publicly available at this time but may be obtained from the authors upon reasonable request.

\end{backmatter}

\bibliography{sample}

\end{document}